\documentclass{article}

\usepackage{arxiv}

\usepackage[ruled,vlined]{algorithm2e}
\usepackage{amsmath}
\usepackage{caption}
\usepackage{float}
\usepackage{graphicx}
\usepackage{subcaption}
\usepackage{multirow}
\usepackage{natbib}
\usepackage{hyperref}
\usepackage{url}

\title{Incremental Learning for Personalized Recommender Systems}

\author{ {Yunbo Ouyang, Jun Shi, Haichao Wei, Huiji Gao} \\
	LinkedIn Corporation\\
	Mountain View, CA 94043 \\
	\texttt{youyang, jshi, hawei, hgao@linkedin.com} \\
}

\renewcommand{\shorttitle}{Incremental Learning}

\begin{document}

\maketitle

\begin{abstract}
Ubiquitous personalized recommender systems are built to achieve two seemingly conflicting goals, to serve high quality content tailored to individual user's taste and to adapt quickly to the ever changing environment. The former requires a complex machine learning model that is trained on a large amount of data; the latter requires frequent update to the model. We present an incremental learning solution to provide both the training efficiency and the model quality. Our solution is based on sequential Bayesian update and quadratic approximation. Our focus is on large-scale personalized logistic regression models, with extensions to deep learning models. This paper fills in the gap between the theory and the practice by addressing a few implementation challenges that arise when applying incremental learning to large personalized recommender systems. Detailed offline and online experiments demonstrated our approach can significantly shorten the training time while maintaining the model accuracy. The solution is deployed in LinkedIn and directly applicable to industrial scale recommender systems. 
\end{abstract}

\keywords{incremental learning, large-scale recommender systems, personalization}

\section{Introduction}
Large scale recommender systems deployed face the challenge to adapt to fast evolving eco-systems. New content such as Ads, news feed, job postings comes in every second, which can only be captured if we dynamically refine machine learning models frequently. In order to keep up with the latest trend, it is tempting to take the last trained model as an initial start point, then continue to train it on the latest dataset, known as warm start in this paper. However, the model tends to overfit on the new dataset and forget what has been learned so far. This phenomenon is often referred to as catastrophic forgetting \citep{goodfellow2013empirical}. 

This problem can be mitigated if we train models on both the past and the latest datasets. This however results in the ever growing training time which conflicts with the goal of a fast model refreshing rate. In this paper, we present an incremental learning approach to address both aspects of the model training. (1) Model training speed: since we only need to train on the incremental portion of data, faster training speed is achieved; (2) Model quality: key quantities from the previous training are used to construct an informative prior,  so the model remembers the essential information from the past data and learn new information from the latest training data.

Accelerating the model training and maintaining the model quality are challenging and sometimes conflicting with each other when applying incremental learning in large-scale personalized recommender systems. In this paper we proposed an incremental learning method with different levels of approximation to carefully balance the model training speed and the model quality. When deploying our solution, we examined different factors which can significantly influence the model quality and the storage cost. 

The proposed incremental learning method applies to generic machine learning models. First we apply the methodology on large-scale personalized logistic regression models due to their theoretical simplicity and model training efficiency. In particular, the experiments are conducted in the framework of Generalized Linear Mixed effects models (GLMix) \citep{zhang:16}, the core of LinkedIn personalized recommender systems. The lessons learned from GLMix are applied to a deep learning model for video recommendation. Similar training speed improvement is observed.

\section{Related Work}
Classic machine learning algorithms are mostly based on batch learning, i.e., the model is trained on a fixed dataset, then deployed for online inference. The assumption is that one can afford to wait until the data is accumulated before a new model is trained. This paradigm does not apply well to data streams where the model needs to be updated before a full dataset is available. Incremental learning fills the gap by learning from the few newly available samples without resorting to a full training. Similar concepts such as online learning, continual learning \citep{Gepperth:16, parisi2019continual, de2019continual} appear in different contexts in the literature.

\citet{schlimmer:86} discussed a case of incremental concept induction.  \citet{alche:01} proposed local strategies to train SVM incrementally. In their review paper, \citet{Gepperth:16} detailed the challenges faced by incremental learning, and gave an overview about popular approaches, its theoretical foundations, and recent applications. Their discussions however do not apply to recommender systems directly.

In the context of artificial intelligence, incremental learning is sometimes referred to as continual learning. The goal is to teach the model to continuously learn from new tasks. Deep learning models often suffer catastrophic forgetting, i.e., doing well on the new task but suffering a significant performance drop on the old tasks. Methods to overcome this shortcoming includes Elastic Weight Consolidation (EWC) \citep{Kirkpatrick:16}, Learning without Forgetting (LwF) \citep{li2017learning}, Gradient Episodic Memory (GEM) \citep{Lopez:17}, Dynamically Expandable Networks (DEN) \citep{Yoon:18}, Averaged Gradient Episodic Memory (A-GEM) \citep{chaudhry2018efficient}. LwF and DEN put constraints on network architecture, which are not suitable for GLMix. GEM and A-GEM reserve past samples so they add additional operational cost. EWC \citep{Kirkpatrick:16} is similar to our approach in one specific case. However EWC has to use another layer of approximation to avoid computing the second order derivatives in deep learning. Such approximation is not suitable for GLMix. We will detail the connection in a later section.

Under the framework of linear models, \citet{criteo-talk} applied incremental learning to Ads click-through-rate prediction on generalized linear models. \citet{ramanath2020lambda} proposed Lambda Learner for Ads recommendation. Our work differs in two aspects. First, we proposed and evaluated different approaches to approximate the Hessian matrix in large scale personalized recommender systems, allowing a fine-grained trade-off between the complexity and the performance. Second, we extended the above methodology on deep learning models.

\section{Personalized Recommender Systems}
To improve users' experience, many recommender systems are designed with personalization capabilities. Personalized recommender systems generate content which fits every member's intrinsically different interest. The major challenge to build a scalable personalized recommender system is to tackle high dimensional entity identifier (Id) features. Although deep learning models can generate embedding features for sparse Id features, the training is time consuming and the architecture is complicated. Generalized Linear Mixed Effects model (GLMix) decomposes a personalized recommender system into 2 submodels: one fixed effects model to capture the general trend which is invariant to different entities, and random effects models to build individual linear models for different types of sparse Id features. Compared with deep learning models, GLMix trains faster because it incorporates a divide-and-conquer approach: we first train the fixed effects model, and then train random effects models on the residuals after scoring the fixed effects model, and go back to fixed effects model training again until convergence \citep{zhang:16}. Although we focus on incremental learning on GLMix in this paper, the theory is readily extended to deep learning recommender systems with minor modifications. See Section \ref{sec:extention-dl} for an example.

\subsection{Generalized Linear Mixed Effects Models}
A mixed effects model consists of a fixed effects model and random effects models. The former captures the overall trend, which is trained on the entire dataset. The latter reflects individuality since it is trained on samples pertaining to each individual entity. In GLMix models, both the fixed effects model and random effects models are linear models. In the case of logistic regression, the overall score (logit) of a GLMix model is the sum of scores from the constituent fixed effects model and random effects models. \citet{zhang:16} used GLMix models for personalized recommendation. Take LinkedIn job recommendation as an example. In $n$-th observation, member $i$ interacts with job $j$. The total score $s_n$ is
\begin{equation} \label{eq:glmix}
    \mathbf{s}_n = \mathbf{x}_{\text{fixed}, n}^T\mathbf{\beta}_{\text{fixed}} + 
    \mathbf{z}_{\text{per-member},n}^T\mathbf{\beta}_{\text{per-member}, i} +
    \mathbf{z}_{\text{per-job},n}^T\mathbf{\beta}_{\text{per-job}, j}.
\end{equation}
The first term on the right hand side is from the fixed effects model. The second and the third term represent the contribution of per-member and per-job random effects models. $\mathbf{x}_{\text{fixed}, n}$, $\mathbf{z}_{\text{per-member},n}$, $\mathbf{z}_{\text{per-job},n}$ are feature vectors for the fixed effects model, the per-member and the per-job random effects model respectively. $\mathbf{\beta}_{\text{fixed}}$, $\mathbf{\beta}_{\text{per-member}, i}$, $\mathbf{\beta}_{\text{per-job}, j}$ are coefficients vectors for three types of models respectively. 

Since random effects models represent the entity propensity, they should be updated more frequently than the fixed effects model to capture the propensity change. In practice we update the fixed effects model in a much lower frequency (e.g., update monthly) than random effects models (e.g., update daily). The random effects model training often takes 6 - 12 hours because the training data contains millions of entities with high dimensional features. Applying incremental learning in the random effects model training is desired to promptly detect the change. In the next section, we will go through the methodology that enables us to tremendously increase the update frequency.

\section{Methodology}
A recommender system can be denoted as $y=f(\boldsymbol{x}, \boldsymbol{w})$ where $\boldsymbol{x}$ is the input feature vector, $\boldsymbol{w}$ is the model weight vector and $y$ is the label. To simplify the presentation, we consider binary labels where $y$ is either 0 or 1. The principles learned from the binary setup can be seamlessly extended to more complex multi-class labels.

\begin{figure}[h]
     \centering
     \begin{subfigure}[a]{0.5\textwidth}
         \centering
         \includegraphics[trim=0 80 0 50,clip,width=\textwidth]{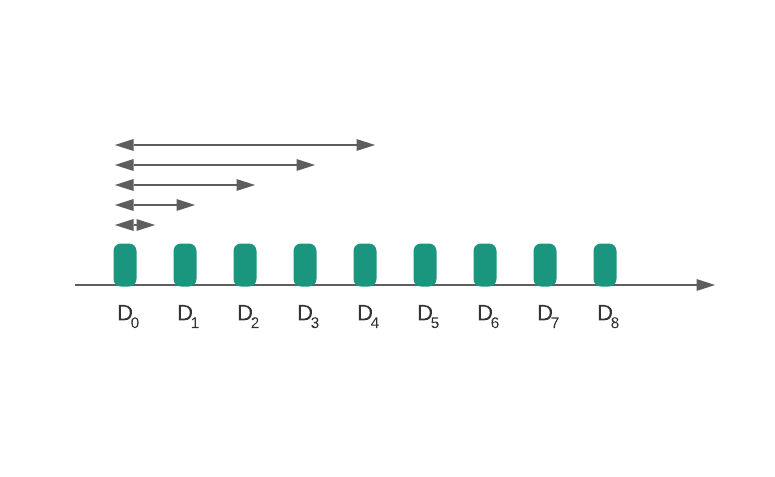}
         \caption{Train on the entire history.}
         \label{fig:growing_window}
     \end{subfigure}
     \begin{subfigure}[b]{0.5\textwidth}
         \centering
         \includegraphics[trim=0 10 0 10,clip,width=\textwidth]{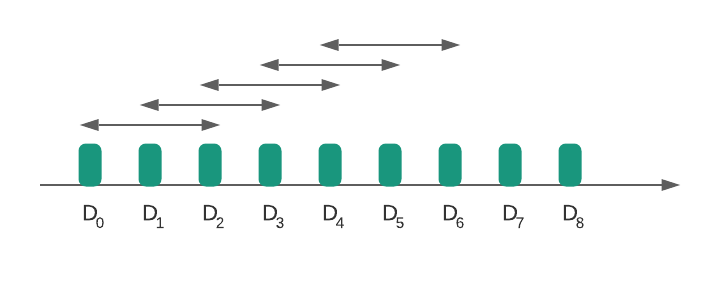}
         \caption{Train on a fixed-length sliding window.}
         \label{fig:sliding_window}
     \end{subfigure}
        \caption{Streaming datasets and training strategies}
        \label{fig:streaming_datasets}
\end{figure}

Suppose there is a stream of datasets $\{\mathcal{D}_0, \mathcal{D}_1, ...\}$ indexed by time. At time $t$, we have access to $\mathcal{D}_0^t=\{\mathcal{D}_0, \mathcal{D}_1, ..., \mathcal{D}_t\}$. Our goal is to build a model based on the stream. As illustrated in Figure \ref{fig:streaming_datasets}(\subref{fig:growing_window}), one approach is to train the model on the entire history $\mathcal{D}_0^t$. The training datasets quickly grow overwhelmingly large. To constrain the training dataset size, we can place a fixed length sliding window on the stream, only the datasets inside the window are used for training as in Figure \ref{fig:streaming_datasets}(\subref{fig:sliding_window}). For example, if at time $t$ we choose window size $l$ to train the model, we are using $\mathcal{D}_{t-l+1}^t=\{\mathcal{D}_{t-l+1}, \mathcal{D}_{t-l+2}, ..., \mathcal{D}_t\}$. This approach re-trains the entire model even if there is only one more new dataset, which does not scale well with a large window length. Another approach is to use a small window length for retraining but initialize the model coefficients based on the model trained previously. For example, we set $l = 1$ and use the model at $t-1$ to initialize the model training at time $t$. This approach, named as warm start, can improve the training efficiency but may result in a model that forgets what is learned in the past. In what follows, we introduce an approach that maintains both the training efficiency and the knowledge learned from the past.

\subsection{Cold Start, Warm Start and Incremental Learning}
Before we dive deep into incremental learning, let us distinguish \textit{cold start}, \textit{warm start} and \textit{incremental learning}. The meaning of the three terms varies in different context. Here we define cold start at $t$ as training on a subset of $\mathcal{D}_{0}^t=\{\mathcal{D}_0, \cdots, \mathcal{D}_t\}$ without use of any previous model. Warm start at $t$ means we only train on $\mathcal{D}_t$ and initialize the coefficients with the prior model trained at $t-1$. Incremental learning means we train only on $\mathcal{D}_t$ and utilizes the model trained at $t-1$, but in a more sophisticated way than just treating it as the initial weights. In Section \ref{sec:bayesian-update} and \ref{sec:quadratic-approx} we explain incremental learning from two different perspectives.

\subsection{Sequential Bayesian Update} \label{sec:bayesian-update}
Incremental learning can be formulated as a sequential Bayesian update problem \citep{bishop-prml, murphy-mlapr, sarkka2013bayesian} (also known as Bayes filter or recursive Bayesian estimation). Our goal is to update the distribution of the model weights sequentially as the new training samples are collected. Denote all the datasets up to $t$ as $\mathcal{D}_0^{t}$ and the current dataset as $\mathcal{D}_t$. According to Bayesian principle, the \emph{posterior} distribution of the weights given $\mathcal{D}_0^{t}$ is (in log domain):
\begin{equation}
\begin{split}
\log\text{Pr}(\mathbf{w}|\mathcal{D}^{t}_0) =& \log\text{Pr}(\mathbf{w}|\mathcal{D}_t, \mathcal{D}^{t-1}_0) \\
=& \log\text{Pr}(\mathbf{w}|\mathcal{D}^{t-1}_0) + \log\text{Pr}(\mathcal{D}_t|\mathbf{w}, \mathcal{D}^{t-1}_0) - \log\text{Pr}(\mathcal{D}_t|\mathcal{D}^{t-1}_0)\\
=& \log\text{Pr}(\mathbf{w}|\mathcal{D}^{t-1}_0) + \log\text{Pr}(\mathcal{D}_t|\mathbf{w})+\text{Constant}\\
=& \text{(i)} + \text{(ii)} + \text{(iii)}.
\end{split}
\end{equation}
The interpretation of the above 3 terms is as follows:
\begin{itemize}
    \item (i): The posterior distribution given the dataset up to time $t-1$, which is then used as the prior for computing $\log\text{Pr}(\mathbf{w}|\mathcal{D}^{t}_0)$.
    \item (ii): The probability $\text{Pr}(\mathcal{D}_t|\mathbf{w}, \mathcal{D}^{t-1}_0)=\text{Pr}(\mathcal{D}_t|\mathbf{w})$ is the likelihood of the dataset $\mathcal{D}_t$ given the weight $\mathbf{w}$.
    \item (iii): The last term  $-\log\text{Pr}(\mathcal{D}_t|\mathcal{D}^{t-1}_0)$ can be dropped from optimization since $\mathbf{w}$ is the variable to optimize and (iii) is unrelated to $\mathbf{w}$.
\end{itemize}

Term (i) usually does not have a closed-form. However, under some mild conditions (i) can be approximated by Gaussian distribution as below (This approximation is known as Laplacian Approximation \citep{de1981asymptotic}):
\begin{equation} \label{eq:laplacian}
    \log\text{Pr}(\mathbf{w}|\mathcal{D}^{t-1}_0) \approx -\frac{1}{2}(\mathbf{w}-\mathbf{w}_{t-1})^T\Sigma^{-1}_{t-1}(\mathbf{w}-\mathbf{w}_{t-1}) + \text{Constant}.
\end{equation}
where $\mathbf{w}_{t-1}$ and $\Sigma_{t-1}$ are the posterior mean and the posterior covariance matrix of $\mathbf{w}$. The approximation in (\ref{eq:laplacian}) serves as a regularization term which pushes the new model to remember the prior model, avoiding catastrophic forgetting. This fundamentally improves warm start since in every optimization step this regularization term will be taken into consideration. 

For logistic regression case, the computation of the covariance matrix $\Sigma_{t-1}$ is documented in Page 119 of \citep{mccullagh1989generalized}. Given features $\mathbf{x_i}$ and label $y_i$, the log likelihood, $\log\text{Pr}(\mathcal{D}_t|\mathbf{w})$, of logistic regression is 
\begin{equation} \label{eq:likelihood}
    \log\text{Pr}(\mathcal{D}_t|\mathbf{w}) = \sum_{(\mathbf{x_i},y_i)\in\mathcal{D}_t }[-\log(1+e^{-\mathbf{w}^T\mathbf{x}_i})\cdot y_i-\log(1+e^{\mathbf{w}^T\mathbf{x}_i})\cdot(1-y_i)].
\end{equation}

Combining (\ref{eq:laplacian}) and (\ref{eq:likelihood}), we arrive at the objective function according to the sequential Bayesian update rule, with the forgetting factor $\lambda_f$ to adjust the importance of the prior distribution.
\begin{equation} \label{eq:sbu}
\begin{split}
\log\text{Pr}(\mathbf{w}|\mathcal{D}^{t}_0) = & \sum_{(\mathbf{x_i},y_i)\in\mathcal{D}_t }[-\log(1+e^{-\mathbf{w}^T\mathbf{x}_i})\cdot y_i-\log(1+e^{\mathbf{w}^T\mathbf{x}_i})\cdot(1-y_i)] \\
     &+[-\lambda_f/2\times(\mathbf{w}-\mathbf{w}_{t-1})^T\Sigma^{-1}_{t-1}(\mathbf{w}-\mathbf{w}_{t-1})] + \text{Constant}.
\end{split}
\end{equation}
The hyper-parameter $\lambda_f$ is the forgetting factor. Our offline experiments suggest the performance is the best when we set $\lambda_f$ around 1. $\lambda_f$  controls how much weight we should assign to the prior model $\log\text{Pr}(\mathbf{w}|\mathcal{D}^{t-1}_0)$.

\subsection{Quadratic Approximation}  \label{sec:quadratic-approx}
Sequential Bayesian update applies to not only logistic regression, but also a large family of probability distributions. For a generic loss function which does not have probabilistic interpretation, we can use quadratic approximation instead. In fact, (\ref{eq:sbu}) can be derived from the quadratic approximation of the loss function for the logistic regression case. The total loss at time $t$ can be split into the contribution from $\mathcal{D}_t$ and $\mathcal{D}^{t-1}_0$ as follows:
\begin{equation}
    \text{total\_loss}(\mathbf{w}) = \sum_{i \in \mathcal{D}_t}\text{loss}_i(\mathbf{w}) + \sum_{i \in \mathcal{D}_0^{t-1}}\text{loss}_i(\mathbf{w}).
\end{equation}

If we expand the second term on the right hand side around the last estimated weight vector $\mathbf{w}_{t-1}$, we obtain the following (this is the second-order Taylor expansion when $\lambda_f = 1$):

\begin{equation} \label{eq:hessian}
    \sum_{i \in \mathcal{D}_0^{t-1}}\text{loss}_i(\mathbf{w}) \approx \sum_{i \in \mathcal{D}_0^{t-1}}\text{loss}_i(\mathbf{w}_{t-1}) + \lambda_f/2\times(\mathbf{w}-\mathbf{w}_{t-1})^T\mathcal{H}_{t-1}(\mathbf{w}-\mathbf{w}_{t-1}).
\end{equation}
where $\mathcal{H}_{t-1}$ is the Hessian matrix with regard to $\mathbf{w}_{t-1}$ over $\mathcal{D}_0^{t-1}$. The first order term vanishes since $\mathbf{w}_{t-1}$ achieves the minimal loss on $\mathcal{D}_0^{t-1}$, resulting in zero gradient. $\lambda_f$ is again the forgetting factor for adjusting the contribution from the past samples. 

Comparing (\ref{eq:sbu}) with (\ref{eq:hessian}) we conclude $\Sigma^{-1}_{t-1} = \mathcal{H}_{t-1}$. Sequential Bayesian update and quadratic approximation are equivalent. 

\subsection{Hessian Approximation in GLMix}\label{sec:approglmix}
In GLMix we build one model for one entity (e.g., one member), resulting in a tremendous number of models. The number of features $p$ of one entity's model is also large. The computational and the storage cost for computing $\mathcal{H}$ for one entity is $O(p^2)$. To reduce the cost we propose two approximation methods as follows.

\subsubsection{Diagonal Approximation}\label{sec:diagappro}

Instead of computing the full Hessian matrix $\mathcal{H}$, we only use the diagonal elements $\text{diag}(\mathcal{H})$ in (\ref{eq:hessian}), which significantly reduces the storage and the computational cost to $O(p)$. 

Although the ideas of diagonal approximation and EWC \citep{Kirkpatrick:16} are similar, the approximation levels are different. We explicitly compute the diagonal elements of the Hessian by taking the second order derivatives on the loss function, while EWC works on the sampled empirical Fisher Information Matrix (FIM) \citep{pascanu2013revisiting}, which uses the square of the first order derivatives to avoid computing the Hessian in neural networks. Directly computing Hessian diagonal elements is preferable in GLMix because FIM is only the approximation of them. The extension of diagonal approximation for neural networks is described in Section \ref{sec:adam}.

\subsubsection{DFP Approximation}
Another approach of the Hessian approximation is motivated by quasi-Newton optimization methods. Suppose we have a generic optimization trajectory $x_0, \cdots, x_k$ for a loss function where $x_0$ is the initial point and $x_k$ is the converged point. We apply a Davidon–Fletcher–Powell (DFP) formula \citep{fletcher1963rapidly} to approximate the Hessian by storing the last $m$ optimizing steps $x_k-x_{k-1}, x_{k-1}-x_{k-2}, \cdots, x_{k-m}-x_{k-m-1}$ and their gradient differences $g(x_k)-g(x_{k-1}), g(x_{k-1})-g(x_{k-2}), \cdots, g(x_{k-m})-g(x_{k-m-1})$. $m$ is often referred to as the memory size and in practice we choose $m \leq 10$. Denote
\begin{equation}
\begin{aligned} \label{eq:dfp-notation}
    \Delta x_i&=x_i -x_{i-1}, ~k-m\leq i\leq k,\\
    \Delta g(x_i) &=g(x_i)-g(x_{i-1}), ~k-m\leq i\leq k,\\
    \rho_i &= (\Delta x_i^T\Delta g(x_i))^{-1}, ~k-m\leq i\leq k.
\end{aligned}
\end{equation}

Notice that we do not need $\mathcal{H}_{t-1}$ by itself but the product $\mathcal{H}_{t-1}(\mathbf{w}-\mathbf{w}_{t-1})$ in our optimization engine. We can update the Hessian-vector product without computing the Hessian. Given a vector $d$, we propose Algorithm \ref{al:dfp} to compute the Hessian-vector product. The algorithm is the same as the L-BFGS update algorithm except that we switch the role of $\Delta x_i$ and $\Delta g(x_i) $ \citep{nocedal2006numerical}. The computational and the storage cost for DFP approximation is $O(mp)$. 

\begin{algorithm}[h]
\SetAlgoLined
\KwIn{last $m$ Optimization steps $\Delta x_{k-m}, \cdots, \Delta x_{k}$, last $m$ gradient differences $\Delta g(x_{k-m}), \cdots, \Delta g(x_{k})$, a vector $d$}
\KwOut{Approximation of $\mathcal{H}d$}
 initialization: $r\leftarrow d$\;
 \For{$i$ in $k, \cdots, k-m$ } {
 $\alpha_i\leftarrow \rho_i\Delta g(x_i)^Tr$\;
 $r\leftarrow r-\alpha_i\Delta x_{i}$\;
 }
  $r\leftarrow \frac{\Delta x_{k}^T\Delta g(x_k) }{\Delta x_{k}^T\Delta x_{k}}\times r$\;
 \For{$i$ in $k-m, \cdots, k$ } {
 $\beta \leftarrow \rho_i\Delta x_{i}^Tr$\;
 $r\leftarrow r+(\alpha_i-\beta)\Delta g(x_i)$\;
 }
Return $r$
 \caption{DFP Approximation Algorithm}
 \label{al:dfp}
\end{algorithm}

\subsection{Hessian Approximation in Deep Leaning}\label{sec:adam}

The approaches described in Section \ref{sec:approglmix} cannot be directly extended to deep learning recommendation systems. Unlike GLMix, it is computationally expensive to compute the Hessian (even the diagonal) for deep learning models over the full dataset. The diagonal of the sampled empirical Fisher Information Matrix (FIM) \citep{pascanu2013revisiting, Kirkpatrick:16} is proposed to approximate the diagonal of Hessian. 

However the sampled empirical FIM only captures the local curvature over one point and only uses a small batch of data. As pointed out by \citet{kunstner2019limitations} and \citet{martens2020new}, deep learning loss functions often suffer from pathological curvature, so that FIM evaluated on a single point might not be close to the underlying true curvature. Adaptive Moment Estimation (Adam) \citep{kingma2014adam} mitigates this problem by using a moving average of squared gradients as an approximation to the diagonal of FIM. Following the spirit of Adam, we use Adam's second moment estimation as the approximation of diagonal elements of the Hessian for neural networks. We refer to this method as incremental learning with Adam approximation (Incre Adam) in this paper and will illustrate more in Section \ref{sec:extention-dl}. Incre Adam has close performance compared with cold start and is as fast as warm start.

\section{Implementation}
\subsection{Algorithm Overview}

\begin{algorithm}
\SetAlgoLined
\KwIn{A stream of datasets $\{\mathcal{D}_0, \mathcal{D}_1, \cdots\}$, cold start training frequency $T$}
\KwOut{A stream of model weight vectors $\{\mathbf{w}_0, \mathbf{w}_1, \cdots\}$}
 initialization: $t=0$\;
 \While{True}{
  \eIf{$t \% T == 0$ }{
   cold start on a full history $\mathcal{D}^t_{0}$ or a sliding window history $\mathcal{D}^t_{t-l+1}$ for a window size $l > 1$\;
   compute and store weights $\mathbf{w}_t$ and the (approximate) Hessian $\mathcal{H}_t$ for the next update\;
   }{
   incremental learning on a small dataset $\mathcal{D}_t$\;
   compute and store weights $\mathbf{w}_t$ and the (approximate) Hessian $\mathcal{H}_t$ for the next update\;
  }
  $t \leftarrow t + 1$
 }
 \caption{Model training with incremental learning}
 \label{al:inc}
\end{algorithm}

Our incremental learning algorithm is listed in Algorithm \ref{al:inc}. Periodical cold start is required to avoid the performance degradation. The cold start training is conducted on large datasets with weeks' or months' samples. The subsequent incremental learning updates are on small datasets with day's or even hour's samples. At the end of each training the weights and the (approximate) Hessian is computed and fed to the next update. 

\subsection{Photon-ML Based Implementation}
\begin{figure*}[h]
     \centering
     \includegraphics[width=155mm]{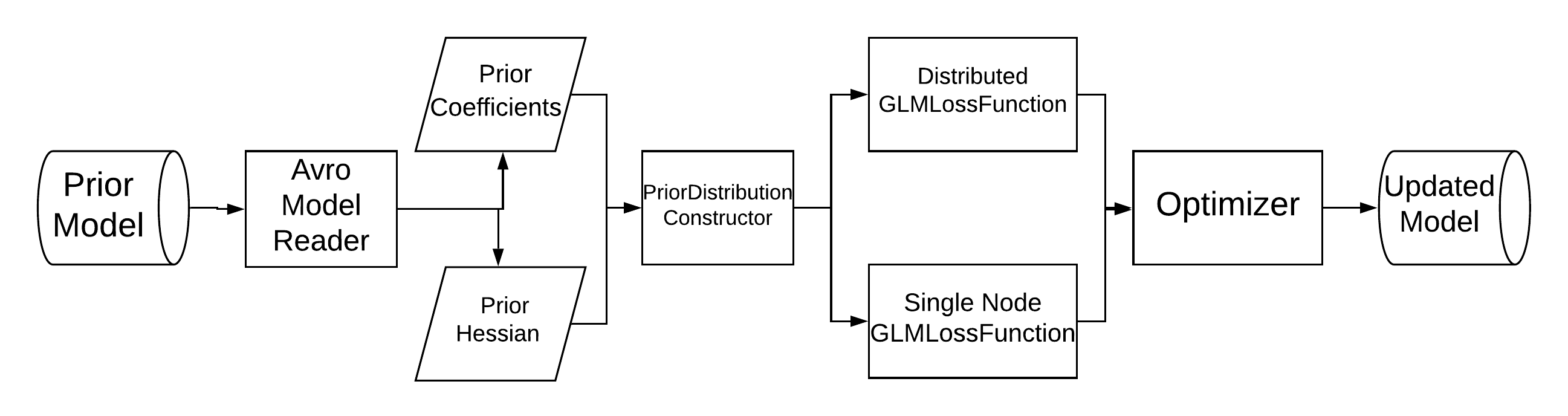}
     \caption{Photon-ML Incremental Learning Implementation Diagram}
     \label{fig:photon-ml}
\vspace{0mm}
\end{figure*}

Our GLMix incremental learning implementation is based on \href{https://github.com/linkedin/photon-ml} {Photon-ML} \citep{photon-ml}. The implementation of deep learning incremental learning is separate and based on TensorFlow.
GLMix incremental learning is roughly divided into the following stages, as illustrated in Figure \ref{fig:photon-ml}.
\begin{itemize}
    \item Load prior models: The prior model is saved in a Hadoop Distributed File System (HDFS) in the avro format. At the beginning of training, it is read distributively to reconstruct the prior coefficients and Hessian matrices. Prior coefficients are constructed in \href{https://github.com/scalanlp/breeze} {Breeze} \citep{breeze}  vector format. Prior Hessians are constructed via different types depending on approximation methods.
    \begin{itemize}
        \item Full: The full Hessian matrix is read as a Breeze matrix.
        \item Diagonal approximation: A Breeze vector is used to store Hessian diagonals. 
        \item DFP approximation: Two sequences of Breeze vectors are constructed: one for optimization steps $\Delta x_i$ and another one for gradient differences $\Delta g(x_i)$. 
    \end{itemize}
    
    Both prior coefficients and Hessians are indexed by entity Ids for random effects models. The number of entities can be more than 10 millions.
    
    \item Add prior models to loss functions: First we construct a PriorDistribution object which bundles prior coefficients and the prior (Approximate) Hessian for each entity Id. After the training data is loaded and distributed to executors according to entity Ids, PriorDistributions are joined with training the data by entity Ids. PriorDistributions output regularization values and their gradients in the optimization step. 
    \item Train a GLMix model: We combine PriorDistributions and training data to construct loss functions, which differs for different model types. DistributedGLMLossFunction targets on training fixed effects models while SingleNodeGLMLossFunction is responsible for training random effects models. Loss functions are sent to an optimizer to minimize. L-BFGS \citep{nocedal2006numerical}, Trust Region New method \citep{sorensen1982newton} and Orthant-Wise Limited-memory Quasi-Newton method \citep{schmidt2009optimizing} are supported optimizers in Photon-ML. After convergence, coefficients and the (approximate) Hessian matrices are stored in HDFS with entity Ids as indices.
\end{itemize}

\section{Offline Experiments}
We conducted offline experiments on MovieLens-20M dataset \citep{movielens20m}, Criteo 1TB Click dataset \citep{criteotb}, and two LinkedIn job recommendation datasets, with the focus on GLMix in this section. The time granularity is different across the four: MovieLens-20M dataset contains 20-year data with many low-engaged users; Criteo dataset only has daily data; hourly data is available for LinkedIn job recommendation datasets. Therefore we choose different experiment settings to be introduced in detail. We apply incremental learning on both fixed and random effects models on MovieLens-20M and Criteo datasets. However, to be consistent with the actual production model in LinkedIn, we apply incremental learning only on random effects models for LinkedIn job recommendation datasets. 

The implementation of incremental learning is available in Photon-ML \citep{photon-ml}. The CPU used is Intel(R) Xeon(R) CPU E5-2683 v3. We repeat our experiments 3 times and report the average results. GLMix models are fitted with the Breeze L-BFGS optimizer with the maximal 100 iterations to guarantee the convergence. The memory size $m = 3$ is chosen for DFP approximation because it can balance the computational cost and the model accuracy well.

\subsection{MovieLens-20M}
\subsubsection{Dataset Description}
We applied the above methodology to \href{https://grouplens.org/datasets/movielens/20m/}{Movielens-20M} dataset. Movielens-20M is a widely used dataset for recommender systems, which contains 20 million ratings from 138,493 users on 27,278 movies. Each user at least has 20 reviews. Each review contains a rating score from 0.5 to 5.0 (0.5 scale).

\subsubsection{Experiment Setup}
We first convert the rating score to a binary label. We define any rating score larger than or equal to 3.0 as a positive rating and any rating score smaller than 3.0 as a negative rating. We build a GLMix model (consists of a fixed effects model and a per-user random effects model) to predict a user’s sentiment to a movie. Each sample contains one user Id, one movie Id, one rating score, one timestamp, and categorical genre features of the movie. Genre features are used in both the fixed effects model and the per-user random effects model. The feature dimension is 2,558.

To mimic the data streaming nature, for each user, we sort his / her reviews by timestamp and evenly split them into 5 temporal phases (we do not split into more than 5 phases in order to reserve at least a few training data points for low-engaged users. The results for splitting into the lower number of phases are similar. Due to the space limit, we omit them). In each temporal phase, we hold out his / her latest review as the test set, the second latest review as the validation set and utilize the remaining data for training.  The validation data is used to tune $\lambda_f$. Grid search is deployed to tune hyperparameters. 

We compare the following 5 methods:
\begin{itemize}
    \item Cold start (Cold): For $t=1, \cdots, 4$, we collect the training data $\mathcal{D}_1^t=\{\mathcal{D}_1, \cdots \mathcal{D}_t\}$, train the model and evaluate on Phase $t + 1$.
    \item Warm start (Warm): For $t=1, \cdots, 4$, we only train on $\mathcal{D}_t$, use the previous model trained at Phase $t - 1$ to initialize the weights and evaluate on Phase $t + 1$.
    \item Incremental learning using the full Hessian (Incre Full): We use the same training data and validation data as Warm in Phase $t$, except we apply Incre Full on each iteration. 
    \item Incremental learning using diagonal approximation (Incre Diag): the same as Incre Full except using diagonal approximation. 
    \item Incremental learning using DFP approximation (Incre DFP): the same as Incre Full except using DFP approximation. 
\end{itemize}

All of 5 methods share the same start point: the cold start model trained on Phase $1$. We report test AUC values from Phase 2 to Phase 5 as well as the training time in minutes. We use 20 Spark executors with 8 GB executor memory for all 5 methods. 

\subsubsection{Experiment Results}
The test AUC and the training time are listed in Table \ref{table:ml20m-auc} and Table \ref{table:ml20m-time}. The best results are marked in bold.
\begin{table}[H]
\centering
\vspace{-2mm}
\begin{tabular}{|c|c|c|c|c|c|}
\hline
        & \multicolumn{1}{c|}{Cold}  & \multicolumn{1}{c|}{Warm}  & \multicolumn{1}{c|}{Incre Diag} & \multicolumn{1}{c|}{Incre Full}     & \multicolumn{1}{c|}{Incre DFP} \\ \hline
Phase 2 & { 0.7234} & { 0.7200}   & { 0.7283}          & { \textbf{0.7307}} & { 0.7288}     \\ \hline
Phase 3 & { 0.7301} & { 0.7251} & { 0.7361}          & { \textbf{0.7393}} & { 0.7382}     \\ \hline
Phase 4 & { 0.7317} & { 0.7254} & { 0.7392}          & { \textbf{0.7422}} & { 0.7414}     \\ \hline
Phase 5 & { 0.7092} & { 0.7069} & { 0.7175}          & { \textbf{0.7208}} & { 0.7193}     \\ \hline
\end{tabular}
\caption{Test AUC for Phase 2 - 5}
\label{table:ml20m-auc}
\vspace{-8mm}
\end{table}

From Table \ref{table:ml20m-auc} it is evident that Incre Full performs the best, followed by Incre DFP and Incre Diag. The performance of all the three incremental learning methods is close, within 0.5\% AUC difference. They all largely outperform cold start and warm start. The reason is that we intentionally throw away higher order terms in the loss function of the past phases' data  via incremental learning. The higher order terms contain a large amount of noise which jeopardizes the performance. Specifically, movie popularity changes year by year in the collected period (20 years) and users tend to review recent movies, which implies the current phase's data is more important than the previous phases' data. Compared with cold start, incremental learning drops some noise of the previous data but keeps the essential information; compared with warm start, incremental learning retains most of the essential information.

We do not see performance degradation after 5 iterations of incremental learning. The results demonstrate incremental learning methods can balance remembering the prior model and learning from the current data. The Hessian preserves the valuable information about the uncertainty of the prior model.

\begin{table}[H]
\centering
\vspace{-2mm}
\begin{tabular}{|c|c|c|c|c|c|}
\hline
        & \multicolumn{1}{c|}{Cold}      & \multicolumn{1}{c|}{Warm}       & \multicolumn{1}{c|}{Incre Diag} & \multicolumn{1}{c|}{Incre Full} & \multicolumn{1}{c|}{Incre DFP} \\ \hline
Phase 2 & {4.3} & { 3.8}  & { \textbf{3.7}}      & { 37.2}          & { 8.6}          \\ \hline
Phase 3 & { 14.9}         & { 8.8}           & { \textbf{4.2}}      & { 42.1}          & { 56.9}         \\ \hline
Phase 4 & { 31.1}         & { \textbf{10.5}} & {10.7}     & { 48.6}          & { 21.3}         \\ \hline
Phase 5 & { 8.4}          & { 13.0}          & { \textbf{4.3}}      & { 63.0}          & { 48.7}         \\ \hline
\end{tabular}
\caption{Training time for Phase 2 - 5 in minutes}
\label{table:ml20m-time}
\vspace{-8mm}
\end{table}

Training time in Table \ref{table:ml20m-time} suggests Incre Diag is the fastest while Incre Full is the slowest. Incre DFP stays in between. The fluctuation in the training time is majorly due to the changing Hadoop YARN cluster environment. 

For a finer granularity, we provide the model loading time (Loading), the model fitting time (Fitting) and the model saving time (Saving) for Phase 2 on one run. Table \ref{table:ml20m-time-breakdown} suggests the bottleneck is the model fitting, which involves the (approximate) Hessian-vector product for diagonal approximation, DFP Approximation and the full Hessian, with the computational cost $O(p), O(mp), O(p^2)$ respectively. The empirical evidence matches the theoretical computational cost.

\begin{table}[H]
\centering
\begin{tabular}{|c|c|c|c|c|c|}
\hline
Stage & Cold & Warm & Incre Diag & Incre DFP & Incre Full \\ \hline
Loading & 0.0   & 0.1  & 0.1        & 0.2       & 0.6        \\ \hline
Fitting & 3.9  & 2.1  & 2.6        & 5.9       & 34.7       \\ \hline
Saving  & 0.2  & 0.1  & 0.1        & 0.2       & 0.1        \\ \hline
\end{tabular}
\caption{Training time breakdown for Phase 2 in minutes}
\label{table:ml20m-time-breakdown}
\vspace{-8mm}
\end{table}

\subsection{Criteo 1TB Click Dataset}
\subsubsection{Dataset Description}
To test our algorithm on the larger recommender system, we apply the above methods to \href{https://ailab.criteo.com/download-criteo-1tb-click-logs-dataset/}{Criteo 1TB Click Dataset}. The dataset consists of 24 days' user click feedback for display Ads. Each sample contains one label, 13 integer features and 26 categorical features. 
\subsubsection{Experiment Setup}
In order to shorten the experiment time, we downsample the negative samples to achieve 1:1 positive-to-negative ratio. The final dataset contains around 12 million samples per day. Following the data processing procedure in \citep{nvidia-deep-learning-examples}, we remove the categorical feature values that appear less than 15 times in the entire dataset. The missing values are padded 0 for the numerical features and "unknown" for the categorical features. It is not obvious which categorical feature is the entity Id since they are anonymized. Judging from the histogram of feature value counts, we select a few features that most likely correspond to entity Ids. The experiment results are similar when any of them is treated as the entity Id. The result reported here is obtained when the first categorical feature (cf1) is used as the entity Id column, which has 26,493 distinct values. The remaining 13 numerical features and 25 categorical features are used to train a GLMix model consisting of a fixed effects model and a per-cf1 random effects model. The feature dimension is around 3.1 million due to the large cardinality of categorical features. The average and the maximal number of nonzero features per entity are 3,000 and 400,000 respectively, causing a storage burden for incremental learning using the full Hessian. Therefore the full Hessian is not used in this experiment. 

\begin{table}[H]
\centering
\vspace{-1mm}
\begin{tabular}{|c|c|c|c|c|}
\hline
              Prediction Day  & \multicolumn{1}{c|}{Cold} & \multicolumn{1}{c|}{Warm} & \multicolumn{1}{c|}{Incre Diag} & \multicolumn{1}{c|}{Incre DFP} \\ \hline
Day 18 & 0.7068  & 0.7082  & 0.7094      & \textbf{0.7101}\\ \hline
Day 19 & 0.7043   & 0.7120    & { \textbf{0.7139}}  & 0.7134\\ \hline
Day 20 & 0.7061                          & { 0.7134}      & { \textbf{0.7162}}  & 0.7147\\ \hline
Day 21 & { 0.7108}      & { 0.7148}      & { \textbf{0.7181}}  & 0.7158\\ \hline
Day 22 & { 0.7083}      & { 0.7167}      & { \textbf{0.7197}} &  0.7176\\ \hline
Day 23 & { 0.7096}      & { 0.7178}      & { \textbf{0.7203}}  & 0.7184\\ \hline
Day 24 & { 0.7119}      & { 0.7211}      & { \textbf{0.7230}}  & 0.7212\\ \hline
\end{tabular}
\caption{Test AUC for Day 18 - 24}
\label{table:criteo-auc}
\vspace{-8mm}
\end{table}

We compare the following 4 methods:
\begin{itemize}
    \item Cold:  At Day 16 we train  on $\mathcal{D}_1^{16}=\{\mathcal{D}_1, \cdots \mathcal{D}_{16}\}$, the dataset from Day 1 to Day 16; in the second iteration at Day 17 we train on $\mathcal{D}_2^{17}$, and so on. We evaluate the model on $\mathcal{D}_{t + 1}$ at day $t$. For instance, at Day 17 we evaluate the model trained on $\mathcal{D}_2^{17}$ on $\mathcal{D}_{18}$. 
    \item Warm: At Day 17 we train on $\mathcal{D}_{17}$ using the cold start model trained from $\mathcal{D}_1^{16}$ to initialize the weights and evaluate on $\mathcal{D}_{18}$. Then we train on $\mathcal{D}_{18}$ using the warm start model trained from $\mathcal{D}_{17}$ to initialize the weights and evaluate on $\mathcal{D}_{19}$, and so on.
    \item Incre Diag: We use the same training datasets, prior models and test datasets as warm start except we apply incremental learning with diagonal approximation on each iteration. 
    \item Incre DFP: This is the same as Incre Diag except we apply DFP approximation.
\end{itemize}
We use 400 Spark executors, each with 48 GB executor memory for Criteo dataset. 

\subsubsection{Experiment Results}
The test AUC and the training time are listed in Table \ref{table:criteo-auc} and Table \ref{table:criteo-time}. The best results are marked in bold.

Table \ref{table:criteo-auc} demonstrates Incre Diag is the best among all four methods. DFP approximation is slightly worse than diagonal approximation. DFP approximation may suffer a  larger deviation from the true Hessian matrix than diagonal approximation for this dataset. Ads data is also time-sensitive. The most recent data is more valuable to predict users' click behavior. Compared with cold start, incremental learning as well as warm start intrinsically down-weights the past data and up-weights the most recent data, which leads to the better performance. We do not see any performance degradation after 7 iterations of incremental learning. Table \ref{table:criteo-time} demonstrates that the training time for cold start doubles compared with other methods. Incremental learning algorithms do not have obvious additional time cost compared with warm start.

\begin{table}[H]
\centering
\begin{tabular}{|c|c|c|c|c|}
\hline
     Prediction Day  & \multicolumn{1}{c|}{Cold} & \multicolumn{1}{c|}{Warm}  & \multicolumn{1}{c|}{Incre Diag} & \multicolumn{1}{c|}{Incre DFP}  \\ \hline
Day 18 & { 233}         & { 149}          & { 125}    &  \textbf{99}\\ \hline
Day 19 & { 228}         & { 143} & { 154}             &  \textbf{89}\\ \hline
Day 20 & { 230}         & { 111} & { 131}             &  \textbf{104}\\ \hline
Day 21 & { 180}         & { \textbf{133}} & { 178}             &  255\\ \hline
Day 22 & { 230}         & { 141}          & { \textbf{129}}    &  133\\ \hline
Day 23 & { 298}         & { 195}          & { 165}    &  \textbf{103}\\ \hline
Day 24 & { 223}         & { \textbf{100}} & { \textbf{100}}    &  114\\ \hline
\end{tabular}
\caption{Training time for Day 18 - 24 in minutes}
\label{table:criteo-time}
\vspace{-8mm}
\end{table}

\subsection{LinkedIn's Job Recommendation Datasets}

We apply incremental learning to LinkedIn job recommendation datasets. We choose 2 datasets: one JYMBII (Jobs You May Be Interested In) dataset and one Jobs Homepage Click dataset. The key differences for these datasets and the experiment setup are summarized as follows:

\begin{itemize}
    \item JYMBII dataset captures members' applying behavior for a given job post from various sources, whereas Jobs Homepage Click dataset captures members' click behavior for promoted job posts in LinkedIn Jobs Homepage Tab. 
    \item The production GLMix model for JYMBII dataset consists of a fixed effects model, a per-member and a per-job random effects model, while the model for Jobs Homepage Click dataset consists of a fixed effects model and a per-job random effects model. In the following experiments we keep the fixed effects model intact. 
    \item To match the production setup, the cold start model for JYMBII dataset uses 75 days' data compared with 14 days for Jobs Homepage Click dataset.
    \item The cold start model training frequencies are different. We retrain the JYMBII model every day while retraining the Jobs Homepage Click model every 12 hours.
    \item The proposed incremental learning training frequencies are every 12 hours for the JYMBII model and every 6 hours for the Jobs Homepage Click model. Incremental learning training frequency is doubled compared with cold start.
\end{itemize}

\subsubsection{JYMBII Dataset}

In what follows, we do not include results for incremental learning using the full Hessian due to the storage concern. We compare the following methods:
\begin{itemize}
    \item Cold: The first cold start model is trained from Day 0, 0:00 to Day 75, 0:00. We denote this dataset as $\mathcal{D}^{\text{75d0h}}_{\text{0d0h}}$.
This first model is evaluated on the next 2 subsequent 12-hour's data, i.e., $\mathcal{D}^{\text{75d12h}}_{\text{75d0h}}$ and $\mathcal{D}^{\text{76d0h}}_{\text{75d12h}}$ respectively. The training data and test data time ranges are shifted one day in the second iteration, and so forth. We repeat the procedure for 5 iterations. 

  \item Warm: We start with the first cold start model and data $\mathcal{D}^{\text{75d12h}}_{\text{75d0h}}$. The model is evaluated on $\mathcal{D}^{\text{76d0h}}_{\text{75d12h}}$. In the second iteration the training data and test data time ranges are shifted 12 hours, and so forth. Every trained model is used as the input in the next iteration. We repeat the procedure 9 times. 
  
  \item Incre Diag, Incre DFP: We use the same training datasets, prior models and test datasets as warm start.
\end{itemize}

\begin{figure}[h]
     \centering
         \includegraphics[width=100mm]{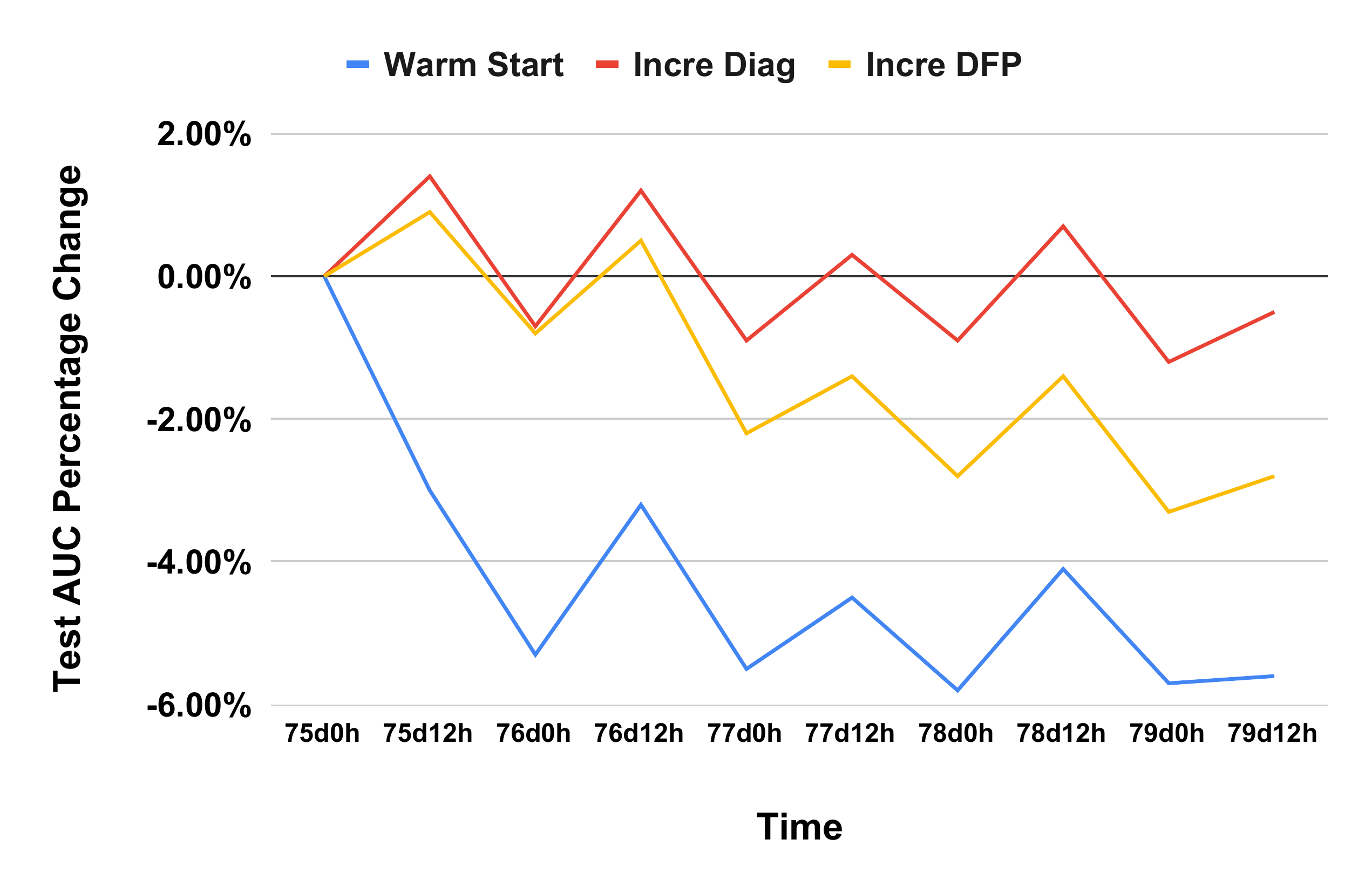}
         \caption{JYMBII Incremental Learning test AUC comparison. Cold start is used as the baseline (0\%). Test AUC percentage change for warm start, incremental learning using diagonal approximation and DFP approximation with respect to cold start is reported. }
         \label{fig:jymbii}
\vspace{-3mm}
\end{figure}

Results are illustrated in Figure \ref{fig:jymbii} and Table \ref{table:jymbii}. We can conclude Incre Diag and cold start has roughly the same performance. Meanwhile, Incre Diag largely outperforms Incre DFP and warm start. Incre DFP underperforms Incre Diag because diagonal elements of the Hessian are dominating in this dataset. DFP approximation may introduce the larger deviation than using diagonal approximation. 

Note that the percentage change of test AUC is higher in 12:00 than that in 0:00. This is because in 12:00 incremental learning can learn from the most recent 12 hours' data (0:00 - 12:00) but cold start cannot use it because of the lower update frequency. Incremental learning also demonstrates the significant time reduction.

\begin{table}[H]
\centering
\begin{tabular}{|c|c|c|}
\hline
           & Averaged Test AUC Change & Total Training Time \\ \hline
Incre Diag & -0.1\%                                 & 24.4 (-41.9\%)              \\ \hline
Incre DFP & -1.3\%                                 & 22.8 (-45.7\%)              \\ \hline
Warm Start & -4.4\%                                 & 17.0 (-59.5\%)              \\ \hline
Cold Start & 0                                      & 42.0               \\ \hline
\end{tabular}
\caption{JYMBII experiment summary statistics. The unit of time is hour. }
\label{table:jymbii}
\end{table}

\subsubsection{Jobs Homepage Click Dataset}

We only include cold start and Incre Diag in this section to match the online A/B test manner: the former is the current production baseline and the latter is the launching candidate because it demonstrates the best performance in the previous JYMBII dataset. 

The experiment setup is very similar as before:
\begin{itemize}
    \item Cold: The first cold start model is trained on $\mathcal{D}^{\text{14d0h}}_{\text{0d0h}}$ and evaluated on the next 2 subsequent 6-hour's data. The training and the test data for the second cold start model is shifted 12 hours, and so forth. We repeat the procedure for 8 iterations.
    
    \item Incre Diag: we start with the first cold start model and incremental data $\mathcal{D}^{\text{14d6h}}_{\text{14d0h}}$. The model is evaluated on $\mathcal{D}^{\text{14d12h}}_{\text{14d6h}}$. In the second iteration we train on $\mathcal{D}^{\text{14d12h}}_{\text{14d6h}}$ and evaluate on $\mathcal{D}_{\text{14d12h}}^{\text{14d18h}}$, and so forth. We repeat the procedure 15 iterations. 
\end{itemize}

\begin{figure}
     \centering
         \includegraphics[width=100mm]{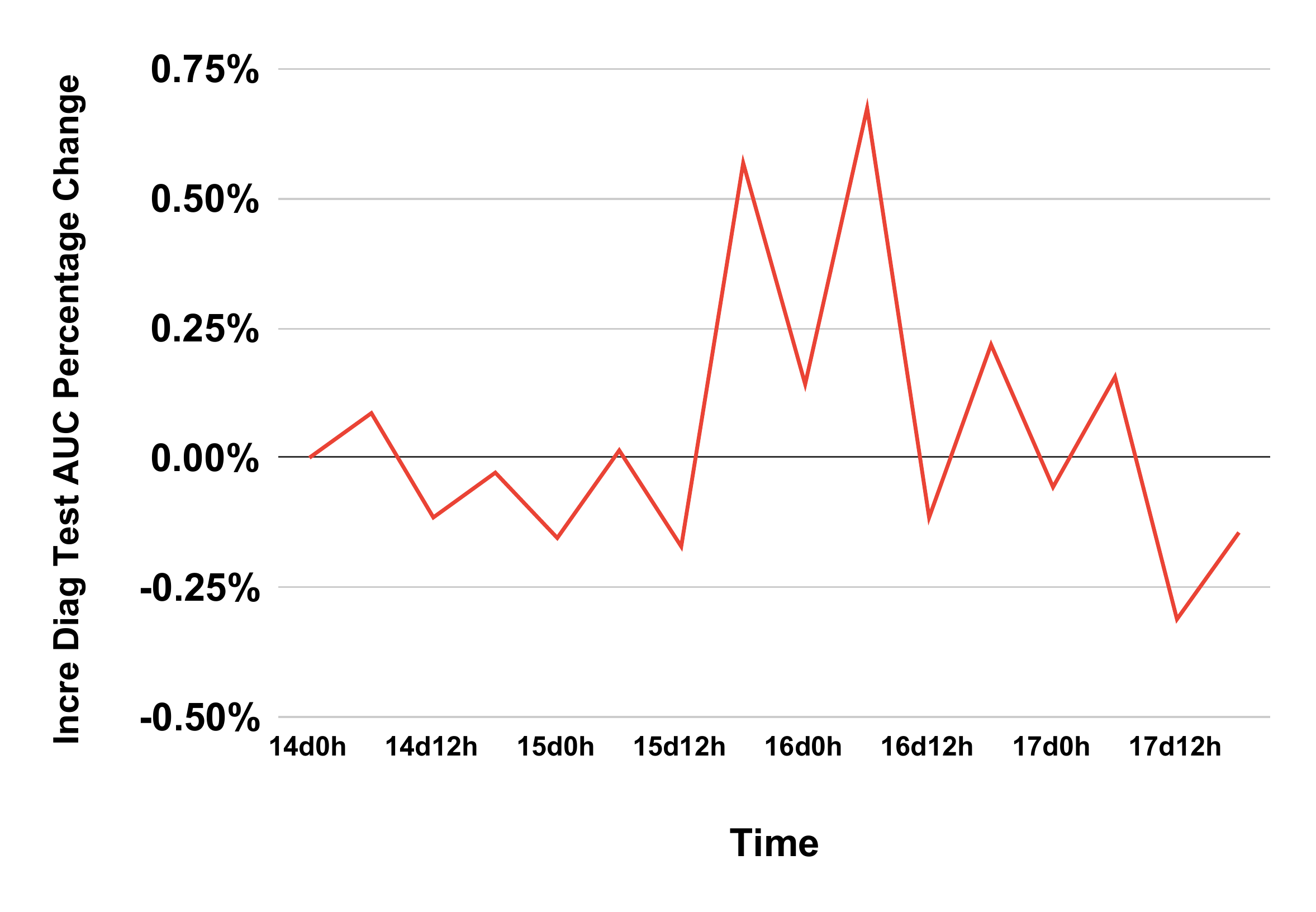}
         \caption{Jobs Homepage click prediction incremental learning model test AUC comparison. Cold start is used as the baseline (0\%). Percentage change for Incre Diag with respect to cold start is reported. }
         \label{fig:jmai}
\vspace{-5mm}
\end{figure}

Table \ref{table:jobs-home-auc} and Figure \ref{fig:jmai} summarize results for cold start and incremental learning. Overall incremental learning slightly boosts the test accuracy (by 0.07\%) and dramatically increase the training efficiency. In the last 2 iterations we see slight performance drop, which suggests the periodical cold start training (model reset) is needed after 15 iterations of incremental learning.

\begin{table}[H]
\centering
\begin{tabular}{|c|c|c|}
\hline
           & Average Test AUC Change & Total Training Time \\ \hline
Incre Diag & 0.1\%                                 & 17.1 (-72.0\%)              \\ \hline
Cold Start & 0                                      & 61.1             \\ \hline
\end{tabular}
\caption{Jobs Homepage click prediction experiment summary statistics. The unit of time is hour.}
\label{table:jobs-home-auc}
\vspace{-3mm}
\end{table}

\section{Online Experiments}
\begin{figure}[h]
     \centering
         \includegraphics[width=85mm]{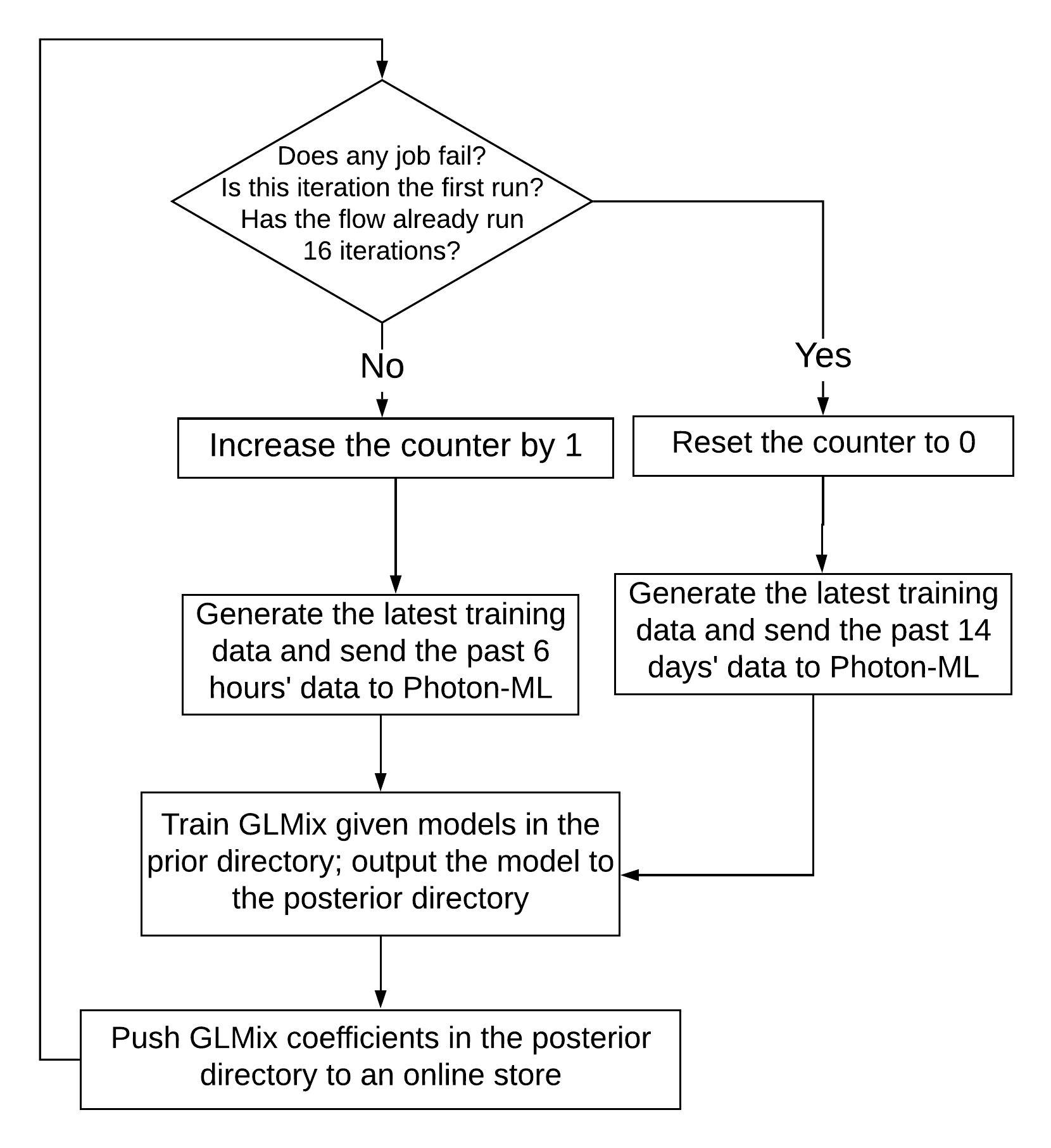}
         \caption{Jobs Homepage Incremental Training Diagram}
         \label{fig:jobs-home-diagram}
\vspace{-2mm}
\end{figure}

We brought the incremental learning model for LinkedIn Jobs Homepage promoted job click prediction in production, as illustrated in Figure \ref{fig:jobs-home-diagram}. The training flow contains three major stages: data preprocessing, Photon-ML training, and model deployment.

In the data preprocessing stage the latest user impression events and click events are fetched. Member features and job features are then joined with these raw events to generate the latest portion of the training data. 

Periodical reset is important. Previous offline experiments suggest cold start should follow after 15 iterations of incremental learning. In this stage we have a switch to determine whether the past 6 hours' data or the past 14 days' data is sent to Photon-ML. A counter is set up to track how many incremental learning iterations the flow has run after cold start. After the counter reaches 16,  the counter is reset to 0 and a cold start training is inserted between the incremental learning updates. This counter also takes failure handling into consideration: the counter is reset to 0 once any job failure is detected. 

In the Photon-ML training stage, we specify one prior directory to store the previous trained model and one posterior directory to store trained model. In the first run, the prior model is copied from the control model. After Photon-ML training, the posterior directory is copied to the prior directory to be used in the next iteration. 

In the model deployment stage the trained model is pushed to a key-value online store with the key as the job Id and the value as the coefficients for online serving. The time spent on this stage is relatively short.

We launched the incremental learning flow with the cold start flow served as the control for A/B testing. The A/B test shows no significant difference with regard to revenue metrics. On the other hand, incremental learning shortens the Photon-ML training time by 71\% and the total flow running time by 31\%, see Table \ref{table:jobs-home-time}. Both results are statistically significant. The saving is majorly due to shortening in the Photon-ML training time.

\begin{table}[H]
\centering
\begin{tabular}{|c|c|c|c|c|}
\hline
                              & Total    & Prep & Training & Deployment \\ \hline
Incre Diag & 6.49 (-31\%) & 4.38              & 1.39 (-71\%)      & 0.72         \\ \hline
Cold       & 9.37         & 3.87              & 4.83              & 0.67        \\ \hline
p-value & 0.0182 & 0.5169 & 0.0001 & 0.7465 \\ \hline
\end{tabular}
\caption{Averaged training hour for incremental training using diagonal approximation and cold start. We select 10 successive iterations and average the training time. The column head "Prep" refers to the data preprocessing stage. "Training" refers to the Photon-ML training stage. "Deployment" refers to the model deployment stage. Reduction in Photon-ML training time and total time is statistically significant.}
\label{table:jobs-home-time}
\vspace{0mm}
\end{table}

\section{Extension to Neural Networks} \label{sec:extention-dl}

\begin{figure}[h]
     \centering
         \includegraphics[width=80mm]{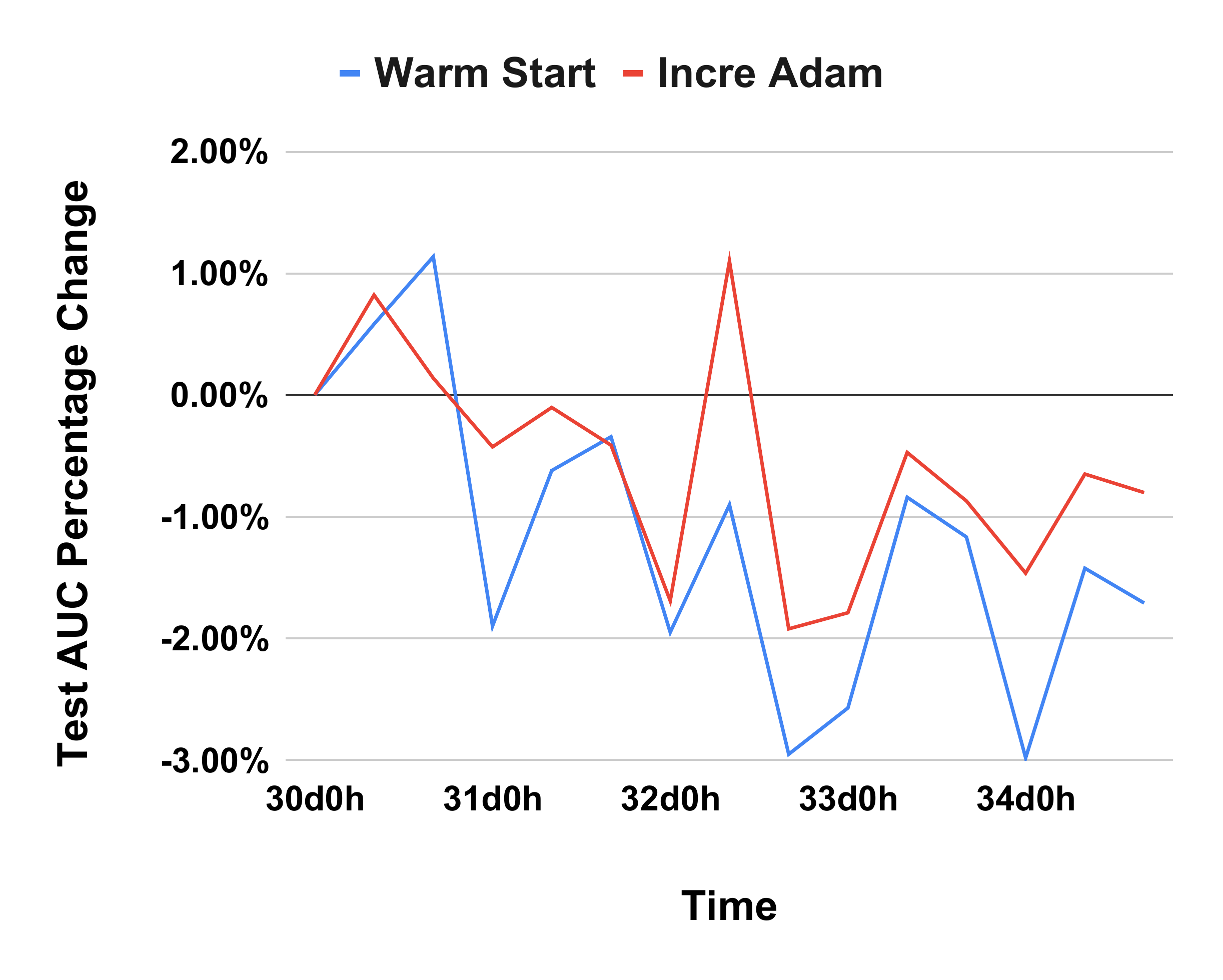}
         \caption{CDML incremental Learning test AUC comparison. Cold start is used as the baseline (0\%). Test AUC percentage change for warm start and incremental learning using Adam approximation with respect to cold start is reported. }
         \label{fig:cdml}
\vspace{-3mm}
\end{figure}

Literature in the neural network incremental learning focuses on applications such as image classification \citep{castro2018end}, face tracking \citep{wu2018deep} and other computer vision tasks. One computer vision application in LinkedIn is the video recommendation in the news feed. Recommendation systems in the news feed relies on video embeddings as features for the downstream ranking. We will focus on the video embedding generation as one application for incremental learning.

Collaborative Deep Metric Learning model \citep{lee2018collaborative} (CDML) is deployed in LinkedIn to generate video embeddings. CDML models the video-video similarity via the member co-watch behavior. The training example in CDML is a triplet which consists of a positive labeled (clicked) video with an unsupervised embedding vector $v_p$, a negative labeled (not clicked by the same member) video with a embedding $v_n$ and an anchor labeled (clicked by the same member) video with a embedding $v_a$. All unsupervised embeddings are pretrained. We are minimizing the following loss
\begin{equation} \label{eq:cdml}
\theta^\ast =\arg\min_{\theta} (\|f_\theta(v_a)-f_\theta(v_p)\|^2 -  \|f_\theta(v_a)-f_\theta(v_n)\|^2 + \alpha, 0),
\end{equation}
where $f$ is a fully connected neural network and $\alpha$ is a pre-specified margin. 

Cold start uses past 30 days' data to retrain CDML models daily. We apply cold start, warm start and incremental learning with Adam approximation (Incre Adam) introduced in Section \ref{sec:adam} to CDML model training:
\begin{itemize}
    \item Cold: At Day 30 we train  on $\mathcal{D}_{0d0h}^{30d0h}$ and evaluate on 3 periods: $\mathcal{D}_{30d0h}^{30d8h}$, $\mathcal{D}_{30d8h}^{30d16h}$ and $\mathcal{D}_{30d16h}^{31d0h}$. In the next iteration the training and the test data is shifted 24 hours.
    \item Warm: we train on $\mathcal{D}_{30d0h}^{30d8h}$ by initializing the weights from the cold start model trained from $\mathcal{D}_{0d0h}^{30d0h}$ and evaluate on $\mathcal{D}_{30d8h}^{30d16h}$. In the next iteration training and test data is shifted 8 hours.
    \item Incre Adam: The training data and the test data is the same as Warm. 
\end{itemize}
To make a fair comparison, for all methods we fix the number of GPUs, the number of epochs, the batch size, the optimization method (Adam), the learning rate and $\alpha$. Test AUC values for the above 3 methods are computed by: (1) generate embeddings $f(v)$ for videos in the database; (2) predict members' click behavior using the embeddings as features in a logistic regression model to compute AUC values.

As demonstrated in Figure \ref{fig:cdml} and Table \ref{table:cdml}, both warm start and incremental learning drastically reduces the training time. Incremental learning underperforms cold start by only 0.6\%, with no increase in the training time compared with warm start. The gap between warm start and cold start is not as large as GLMix. The major reason is that random effects model training is independent for each entity. Once there are limited data for one entity in one time period, that entity's model tends to suffer from catastrophic forgetting. Otherwise in deep learning data for many entities are jointly trained so that catastrophic forgetting is the less severe.

\begin{table}[H]
\centering
\begin{tabular}{|c|c|c|}
\hline
           & Averaged Test AUC Change & Total Training Time \\ \hline
Incre Adam & -0.6\%                                 & 32.7 (-79.0\%)              \\ \hline
Warm Start & -1.2\%                                 & 33.5 (-78.5\%)              \\ \hline
Cold Start & 0                                      & 156.1              \\ \hline
\end{tabular}
\caption{CDML Experiment Summary Statistics. The unit of time is hour.}
\label{table:cdml}
\end{table}

\section{Lessons Learned}
We discuss the following aspects that are crucial to the success of incremental learning. 

\textbf{Cold start frequency}. If we keep incremental learning forever, the model starts to drift away from the original values. This drifting results in performance degradation. Our experiments show periodical cold start is the simplest approach to maintain the model quality. The cold start training frequency $T$ is a hyper-parameter depending on specific use cases. The optimal value can be found by grid search. In our experiment, we found 15 incremental updates per cold start is a good value for LinkedIn Jobs Homepage recommendation.

\textbf{Forgetting factor}. The forgetting factor is used to adjust the contribution of the previous model. For bona fide Bayesian update, the forgetting factor should be fixed at 1.0. In reality, the forgetting factor can be used to filter out the past noise. In addition, when the sample distribution varies significantly between datasets, having a smaller forgetting factor can speed up the model adaptation. If the incremental data is small, a large forgetting factor is desirable to avoid the sudden drift of the models. Our experience shows that 1.0 is a good starting point. The optimal value is usually found between 0.9 and 1.0. The best forgetting factors were fairly close to 1.0 in all our experiments.

\textbf{Hessian approximation}. Full Hessian matrices usually result in the best relevance metrics. However the computational cost and the storage cost can be prohibitively high. Our experience is that diagonal approximation does not result in much metric drop but requires significantly less storage. Hence we use diagonal approximation in the production.

\section{Conclusion}
Incremental learning is a viable solution to reduce the training time while maintaining the model quality. The details such as the forgetting factor, the periodical cold start and the Hessian approximation are crucial to its success in production. Future work is to explore and compare more Hessian approximation methods in deep learning models where the training time places a more severe constraint on the update frequency.

\section*{Acknowledgement}
We would like to thank Keerthi Selvaraj for technical guidance, Qianqi Shen, Aditya Aiyer, Jerry Shen
for assistance in offline/online experiments on LinkedIn Jobs Homepage click prediction, Aman Gupta, Xue Xia and Sirjan Kafle for assistance in CDML video embedding generation.

\newpage

\bibliographystyle{ACM-Reference-Format}
\bibliography{kdd}

\end{document}